# Understanding the origin of magnetocrystalline anisotropy in pure and Fe/Si substituted $SmCo_5$

*Rajiv K. Chouhan[1,2,3], A. K. Pathak[3], and D. Paudyal[3]

[1]CNR-NANO Istituto Nanoscienze, Centro S3, I-41125 Modena, Italy
[2]Harish-Chandra Research Institute, HBNI, Jhunsi, Allahabad 211019, India
[3]The Ames Laboratory, U.S. Department of Energy, Iowa State University, Ames, IA 50011–3020, USA

We report magnetocrystalline anisotropy of pure and Fe/Si substituted $SmCo_5$. The calculations were performed using the advanced density functional theory (DFT) including onsite electron-electron correlation and spin orbit coupling. Si substitution substantially reduces both uniaxial magnetic anisotropy and magnetic moment. Fe substitution with the selective site, on the other hand, enhances the magnetic moment with limited chemical stability. The magnetic hardness of $SmCo_5$ is governed by Sm *4f* localized orbital contributions, which get flatten and split with the substitution of Co (2c) with Si/Fe atoms, except the Fe substitution at 3g site. It is also confirmed that Si substitutions favor the thermodynamic stability on the contrary to diminishing the magnetic and anisotropic effect in $SmCo_5$ at either site.

**INTRODUCTION**

Search for high-performance permanent magnets is always being a challenging task for modern technological applications as they are key driving components for propulsion motors, wind turbines and several other direct or indirect market consumer products. Curie temperature $T_c$, magnetic anisotropy, and saturation magnetization $M_s$, of the materials, are some basic fundamental properties used to classify the permanent magnets. At Curie temperature a material loses its ferromagnetic properties, hence higher the $T_c$, better is the magnets to be used under extreme conditions. Out of all rare earth based permanent magnets, $SmCo_5$ is the champion magnet in terms of uniaxial magneto-crystalline anisotropy energy (MAE) with high Curie temperature. MAE is another important parameter, which defines the hardness with respect to the microscopic characteristics. In spite of the highest uniaxial MAE of ~24.2 $MJ/m^3$ (12.96 meV/f.u.) in $SmCo_5$,[1] the world market is dominated by the $Nd_2Fe_{14}B$ (MAE of ~4.9 $MJ/m^3$) [2,3,4,5,6,7] permanent magnet because of its low energy product. Since after the formulation of $SmCo_5$ in 1970, lots of research has been done to increase the magnetic moment of $SmCo_5$ compared to that of $Nd_2Fe_{14}B$.

The high coercivity of samarium based magnets originates from the Sm sublattice anisotropy, whereas the transition metals such as Co sublattice yields a high curie temperature and thus stabilizes through inter-sublattice exchange.[8] Past studies[9,10] have also confirmed that $T_c$ depends on the interaction of magnetic exchange between adjacent spins. Magnetic anisotropy is the energy required per unit volume to change the orientation of the magnetic moments under the external magnetic field, which is required to achieve the high coercivity in permanent magnets. Magnetic anisotropy also depends on the intrinsic property of a material i.e. does not depend on the microstructure arrangement, where the spin-orbit coupling and its interaction to the crystal electric field created by their neighbor favors the energetically stable magnetic alignment in the specific direction of the lattice. This intrinsic behavior of magnetic anisotropy has the potential to enhance the characteristics of the microstructure within $SmCo_5$. Hence, one can utilize grain boundaries developed through the site substitutions to increase the energy product.[11] Further, the saturation magnetization $M_s$ is the measure of the moment per unit volume defined[12] as maximum energy product $(BH) \max \approx M_s^2$.

---

*rajivchouhan@gmail.com, rajiv.chouhan@nano.cnr.it

In this manuscript, we report the influence of Fe/Si substitutions in the non-equivalent Co (2c) and Co (3g) sites of $SmCo_5$. We have investigated the effect of doping of a *d* block element iron (Fe) and a *p* block element silicon (Si). SQUID magnetometer measurements are performed to investigate the magnetic moment properties of $SmCo_5$. We have also analyzed the adequacy of DFT that includes the onsite *4f* electron correlation and spin orbit coupling calculations, which is very important for investigating the properties of rare earth based permanent magnets.

**METHODS**

We have performed the advanced density functional theory (DFT)[13] based calculations to study electronic, magnetic, and MAE properties of pure and site substituted $SmCo_5$. We have considered Fe and Si substitutions to reduce the Co content and to optimize the magnetocrystalline anisotropy and magnetic moment if possible. We employed full-potential linearized augmented plane wave (FP-LAPW)[14] the method within the framework of generalized gradient approximation, onsite electron correlation, and spin orbit coupling (GGA+U+SOC)[15,16]. To validate our calculations, we have employed a non-collinear plane-wave self-consistent field (PwScf)[17] the method within the GGA+U+SOC framework when needed. We have used the optimized onsite electron correlation parameters U = 6.7 eV and J = 0.7 eV, additionally, it is known in the literature that U= 4 to 7 eV works well for rare earth systems without substantially changing the physical properties[18]. Here, MAE is calculated using the difference between the c-axis and the planar direction total energies.[19] The k-space integrations have been performed with a Brillouin zone mesh measuring at least 13×13×15, which was sufficient for the convergence of total energies ($10^{-6}$Ryd.), charges, and magnetic moments. For $SmCo_5$ the atomic radii for Sm and Co are set as 2.5 and 2.19 with force minimization of 3%. In site-substituted systems, these radii are changed while following the force minimization criteria. As mentioned in the literature[1], slightly higher values of plane-wave cutoff ($RK_{max}$ = 9.0/$R_{MT}^{min}$ = 4.11 a.u.$^{-1}$ and $G_{max}$ = 14$\sqrt{Ry}$) are required for rare earth element systems. Formation energies ($\Delta H_{form}$) of the substitute compounds are calculated by taking the difference of the total energies and chemical potentials of the elements: $\Delta H_{form} = E_{substitution}^{Total} - (\mu_{Sm} + n \Sigma \mu_{element})$, where $E_{substitution}^{Total}$ is the total energy of the substituted compound, *n* is number of atoms per element per f.u., and $\mu_{Sm}$ & $\mu_{element}$ are the chemical potentials of the Sm, and substituted elements (Co, Si, and Fe), respectively. The bulk calculations for pure elements are done with the experimental parameters: Sm[20] (SG = 166, a = 16.97 a.u., alpha = 23.31 degree); Co[21] (SG = P63/mmc, a = 4.738 a.u., c = 7.691 a.u.); Si[22] (SG = 227, a = 10.259 a.u.); Fe[23] (bcc, a = 5.424 a.u.). Finally, to verify the theoretically predicted magnetic moments in SmCo$_5$, we have prepared a sample (as cast) using the arc melting procedure and measured magnetization as a function of the magnetic field using a SQUID magnetometer.

**RESULTS**

SmCo$_5$ forms in the hexagonal CaCu$_5$-type structure (figure 1) with three non-equivalent sites: Sm (1a), Co (2c), and Co (3g). Sm lies in the middle of the hexagonal layer of Co (2c) atoms, and this layer is sandwiched by the plane containing Co (3g) atoms. This structural environment creates a crystal field that splits localized Sm *4f* states and partially quenches the Sm *4f* orbital moment giving rise to a large part of the magnetic anisotropy and net Sm *4f* moment. We have performed volume relaxation by varying the lattice constants c and a. The equilibrium lattice constants came out to be very close to the experimental values (a = 9.45 a.u. and b= 7.48 a.u.)[24] with a volume of 579.19 a.u.$^3$

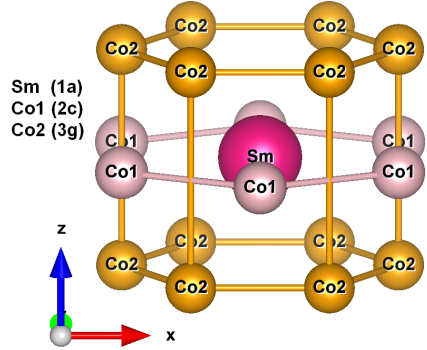

FIG. 1. Crystal structure of SmCo$_5$ representing Co1 (2c site) and Co2 (3g site) of cobalt.

Similar volume relaxation procedure was performed for the Fe, and Si substituted compounds (figure 2). The optimized lattice parameter for SmCo$_2$Fe$_3$, SmFe$_2$Si$_3$, SmCo$_2$Si$_3$, and SmSi$_2$Co$_3$ are mentioned in Table.1. The optimized lattice parameter results indicate that the stability of the Fe substituted structure at 2c, and 3g sites are achieved with the increase in the unit cell volume by 3.9 %, and 5 % in SmCo5, whereas for Si substitution at 2c, and 3g sites the volume of the unit cell change by -1.5 %, and 1.9 %, respectively.

| System | a (in a.u.) | c (in a.u.) | ΔVolume w.r.t SmCo$_5$ |
|---|---|---|---|
| SmFe$_2$Co$_3$ | 9.638 | 7.362 | 3.9 % |
| SmCo$_2$Fe$_3$ | 9.639 | 7.557 | 5 % |
| SmSi$_2$Co$_3$ | 9.92 | 6.725 | -1.5 % |
| SmCo$_2$Si$_3$ | 9.771 | 7.128 | 1.9 % |

Table 1: Lattice parameters for various systems in atomic unit.

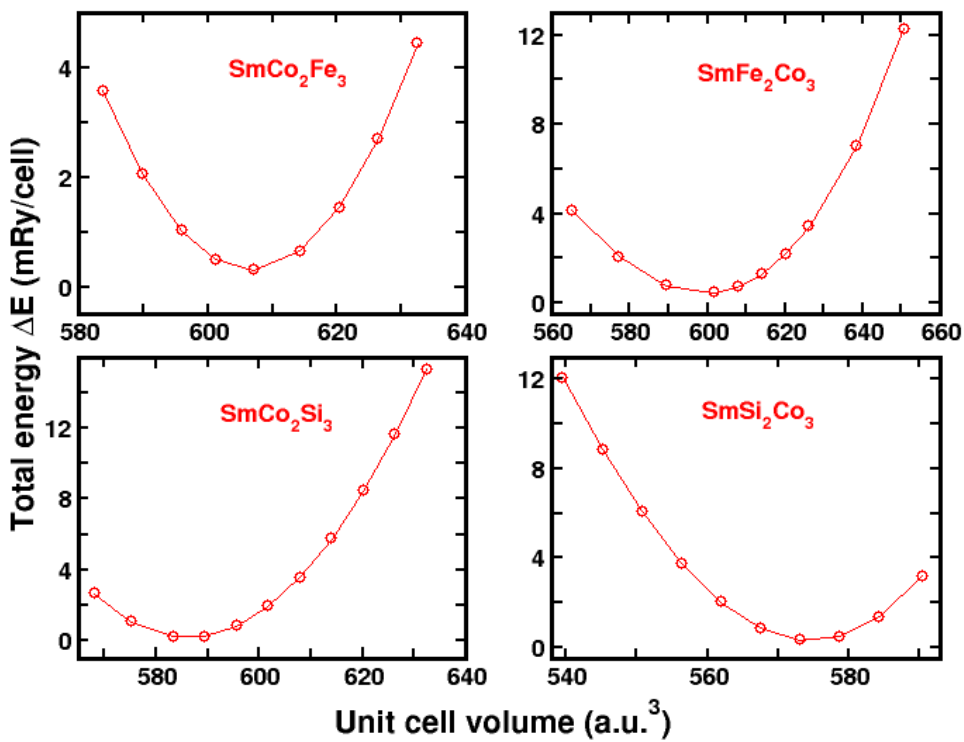

FIG. 2. Volume relaxation of the crystal geometry $SmCo_2(Fe/Si)_3$ and $Sm(Fe/Si)_2Co_3$.

The $\Delta H_{form}$ for the substitute SmCo5 is plotted in figure 3. The data clearly shows that the silicon substitutions are energetically favorable whereas Fe substitution at 2c site is unstable compared to the 3g sites. The instability of iron substitution calculated for iron substitutions are also confirmed by previous results.[25]

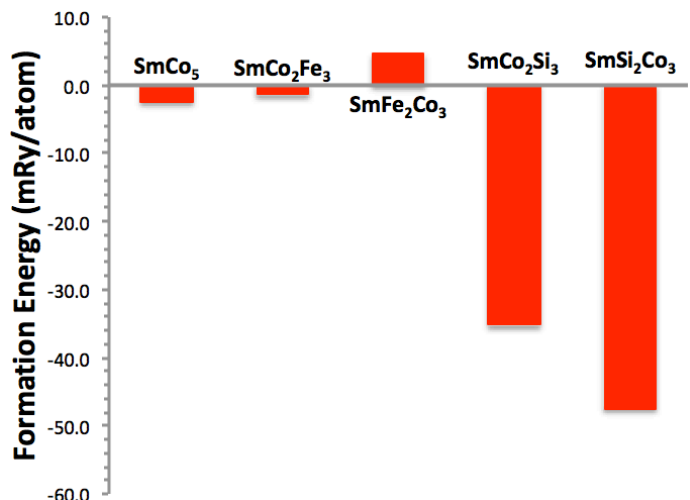

FIG. 3. Formation energies of the $Sm(Si/Fe)_x Co_{(5-x)}$ compounds

The calculated spin magnetic moment (employing GGA+U+SOC) for Sm in $SmCo_5$ is 4.88 $\mu_B$ with the orbital contribution of -1.44 $\mu_B$. The spin moments for Co (2c) (1.55 $\mu_B$) and Co (3g) (1.58 $\mu_B$) are aligned parallel to the spin moment of Sm, and similarly, the orbital moments of Co (2c) (0.14 $\mu_B$) and of Co (3g) (0.12 $\mu_B$) are aligned parallel to the spin moments of Sm. Henceforth, the total magnetic moment comes out to be 11.12 $\mu_B$. This value is large because of the partial quenching of the Sm *4f* orbital moments. The orbital moment also depends heavily on the atomic radius and the value of the onsite Sm *4f* electron correlation. The GGA+SOC calculation shows more negative Sm *4f* orbital moment (-2.29 $\mu_B$), which further reduces the net moment to 10.54 $\mu_B$ which is comparable to the SQUID magnetometer measured the experimental value of ~ 8.02 $\mu_B$ (at 2K) shown in figure 4.

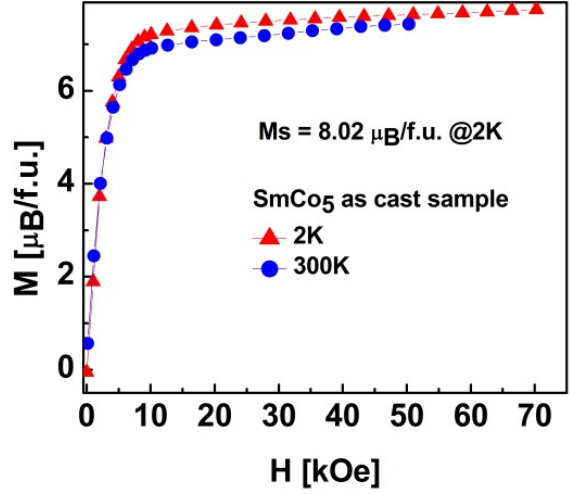

FIG. 4. Magnetization as a function of applied magnetic field at 2k and 300k for bulk SmCo$_5$.

The partially filled Sm *4f* orbitals play a key role in the single ion magnetic anisotropy in SmCo$_5$. The calculated MAE for this compound is 26.06 meV/f.u. (uniaxial) which agrees well with the available experimental values (~13 to 16meV/f.u.).[26,27,28,29,30] The GGA+SOC calculations also show uniaxial MAE of 20.82 meV/f.u., which agrees well with the experimental MAE. There is no sign problem as indicated by earlier studies[31]. Although the MAE with GGA+SOC is much closer to the experiment, however, the occupied and unoccupied *4f* states are incorrectly located at the Fermi level leading to artificial *4f-4f* bonding in SmCo$_5$. Therefore, for better spectroscopy, we need to incorporate Hubbard U. The earlier work using GGA+SOC+U reported a MAE of 21.6 meV/cell,[1] which is smaller than our findings. It is not surprising because the k-points, muffin-tin radii, and Hubbard parameters used in *P. Larson et al.*[31] calculations are different than ours. To reveal the origin of high MAE, we further performed charge density calculations (figure 5) along the uniaxial direction as a side and top view to the plane. It is clearly seen from the charge densities plots that charges are distributed in the uniaxial direction around the Sm atoms and give rise to the magnetocrystalline anisotropy in the system. The origin of axial charge accumulation is because of the hexagonal Co (2c), and planar Co (3g) ring structures (Figures 1, and 5), which bind the Sm atoms between symmetric environments within the planes.

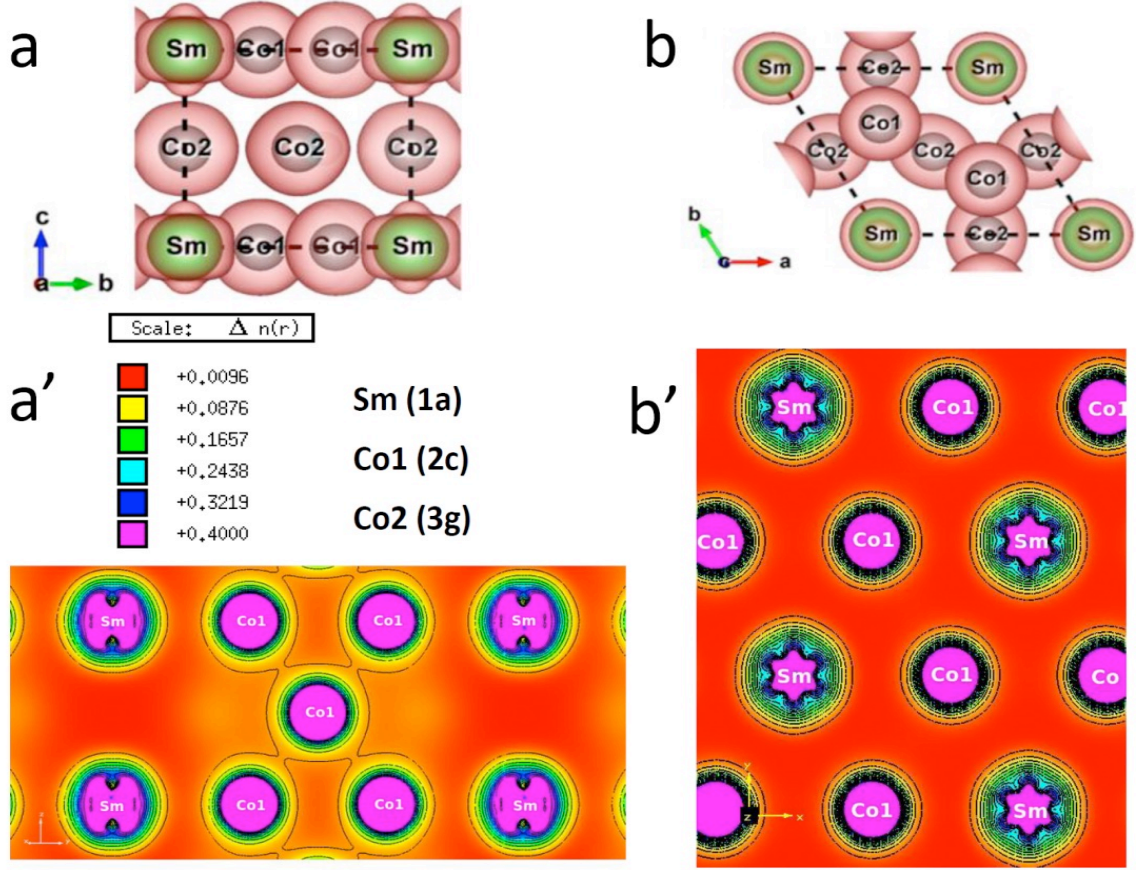

FIG. 5. Charge densities for $SmCo_5$ along c-axis side view (a, a') and c-axis top view (b, b') indicating the uniaxial anisotropy of Sm atoms.

Further, to see the effect of doping in $SmCo_5$ we carried out Fe and Si substitution at Co (2c), and Co (3g) sites. First we doped the constituent atom at the 2c site and found that in $SmFe_2Co_3$ the spin (orbital) moments for Sm (1a), Fe (2c), and Co (3g) sites are 4.90 $\mu_B$ (-2.38 $\mu_B$), 2.60 $\mu_B$ (0.05 $\mu_B$), and 1.54 $\mu_B$ (0.09 $\mu_B$), respectively. The total magnetic moment comes out to be 12.02 $\mu_B$ in $SmFe_2Co_3$. However in the case of Si doping at the 2c site the total magnetic moments drastically dropped to 3.03 $\mu_B$ making a spin contribution of 5.14 $\mu_B$ (orbital -1.39 $\mu_B$), 0.01 $\mu_B$, and 0.37 $\mu_B$ (zero orbital moments) for Sm (1a), Si (2c), and Co (3g), respectively. The calculated MAE values for $SmFe_2Co_3$ and $SmSi_2Co_3$ are 13.05 meV/f.u., and 16.82 meV/f.u., which are along the uniaxial direction similar to that for $SmCo_5$ but with a lower magnitude. However, when we substituted Fe, and Si at the 3g sites the compound has a major doping effect. In the case of $SmCo_2Fe_3$ the spin moments for Sm (1a), Co (2c), and Fe (3g) sites are 4.87 $\mu_B$, 1.60 $\mu_B$, and 2.55 $\mu_B$, whereas the orbital moments are -1.44 $\mu_B$, 0.104 $\mu_B$, and 0.05 $\mu_B$, respectively. This results in the total magnetic moment of 14.02 $\mu_B$ in the $SmCo_2Fe_3$ system. Interestingly, the magnetocrystalline anisotropy for Fe at the 3g site is 28.84 meV/f.u., which is comparable to $SmCo_5$ with a 26% large magnetic moment. But when we substitute Si at the 3g sites the $SmCo_2Si_3$ behaves like a non-magnetic system (total moment ~ 0.6 $\mu_B$) by vanishing the individual magnetic contributions of the Sm, and Co atoms. This is because of the Sm-Co layer is sandwiched and hybridized by the *p* orbital of the newly formed Si layers. This results in the MAE of the entire SmCo$_2$Si$_3$ system to be planar (-0.15 meV/f.u.).

The density of states plot in figure 6 shows the contribution of from the individual atoms in the compounds SmCo$_5$, Sm(Fe/Si)$_2$Co$_3$, and SmCo$_2$(Fe/Si)$_3$. It is clearly visible that the inclusion of onsite correlation in the calculation shifts the Sm (*4f*) DOS to -6.0 eV energy making the split of three states compared to two states shown by GGA+SOC in SmCo$_5$. Also, there is an extra crystal field splitting in one of the *4f* states because of the spin-orbit coupling interaction within the orbitals. The splitting becomes more prominent with the U incorporation in the calculation going from GGA+SOC to GGA+U+SOC. This is one of the main reasons for such high magnetocrystalline anisotropy in SmCo$_5$.

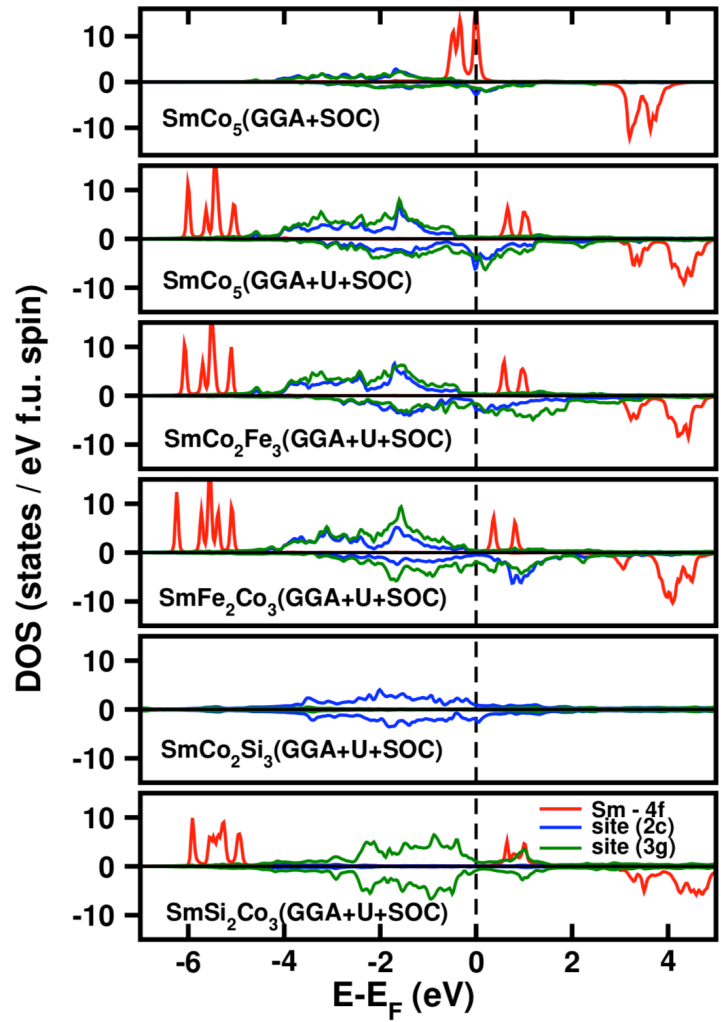

FIG. 6. Individual atomic density of states (DOS) for SmCo$_5$, Sm(Fe/Si)$_2$Co$_3$ and SmCo$_2$(Fe/Si)$_3$. Red line corresponds to the *4f* DOS and blue and green correspond to the DOS of 2c and 3g sites.

The Co (3g) site has a significant contribution to the magnetic moment in $SmCo_5$ and it remains unaffected with the Fe substitution at the 2c site, whereas Si substitution at the 2c site reduces the spin moment contribution of Co (3g) by *p-d* hybridization. It can also be noticed that in the case of $SmFe_2Co_3$, we can see that there is a further splitting of *4f* orbital peak (5 peaks) below the Fermi energy, which may be responsible for lowering of MAE compared to the four peaks in $SmCo_5$. This is not the case in the $SmCo_2Fe_3$ because the anisotropy is comparable to $SmCo_5$ and raising the total magnetic moment contribution by 26% per formula unit. In Si substitution, the magnetic moment and MAE both are affected drastically. At the 3g sites, the Si substitution kills all the contributions of Sm *4f* orbitals. Similarly, in $SmSi_2Co_3$ the Sm *4f* split states are distorted and kill the *3d* spin contribution of Co atoms. As a result, it is suggested that Fe should be substituted at the 3g sites to gain the magnetic moment without compromising the anisotropy of the system.

**Conclusion:**

In conclusion, the substitution mechanism of magnetic Fe and non-magnetic Si atoms suggests that the hexagonal ring of 2c sites containing Sm atom in middle are responsible to maintain the magnetic hardness in the materials. This hardness is defined by the *4f* localized orbital contribution of Sm atoms. This localized orbital get broaden when Co (2c) sites are substituted by the Si, and Fe atoms, which further decrease the anisotropy contribution significantly. Whereas the Co (3g) layer contributes the total magnetic moment of the compound and when it is substituted with Fe atoms, it not only increase the total magnetic moment to 14.02 $\mu_B$ but also boosts the anisotropy by 10% in the $SmCo_2Fe_3$ compared to $SmCo_5$. This PDOS analysis clearly demonstrates that Sm (*4f*) contribution is not affected in $SmCo_2Fe_3$, however, the extra increment of anisotropy comes from the increased PDOS contribution of Fe (3g) layer. Further, we confirm that the Si substitution favors the thermodynamic stability on the contrary to diminishing the magnetic as well as the anisotropic effect of $SmCo_5$ significantly.

**Acknowledgment:**

The research was supported by the Critical Materials Institute, an Energy Innovation Hub funded by the U.S. Department of Energy, Office of Energy Efficiency and Renewable Energy, Advanced Manufacturing Office. This work was performed at the Ames Laboratory, operated for DOE by Iowa State University under Contract No. DE-AC02-07CH11358. R. K. Chouhan also thanks to HPC computational facilities of Harish-Chandra research institute, Prayagraj, India for performing few unfinished calculations. We acknowledge fruitful discussions with Professor Vitalij Pecharsky (group leader), and late Professor Karl A. Gschneidner Jr. (chief scientist, CMI) during the course of this research.